\documentclass[12pt,twoside,pointlessnumbers,smallheadings]{article}
\usepackage{indentfirst}
\input psfig.sty

\newcommand{\BR}{{\cal B}}
\newcommand{\tautorho}{\tau^-\to\pim\piz\nu_{\tau}}
\newcommand{\Btau}{\BR_{\tautorho}}
\newcommand{\BCVC}{\BR^{\rm CVC}_{\tautorho}}
\newcommand{\fpi}{F_{\pi}}

\newcommand{\pim}{\pi^-}
\newcommand{\piz}{\pi^0}
\newcommand{\etap}{\eta^{\prime}}

\newcommand{\psp}{\psi(2S)}

\newcommand{\EE}{e^+e^-}

\newcommand{\pp}{\pi^+\pi^-}

\newcommand{\ppjpsi}{\pi^+\pi^- J/\psi}
\newcommand{\pppsp}{\pi^+\pi^- \psi(2S)}
\newcommand{\beq}{\begin{equation}}
\newcommand{\eeq}{\end{equation}}
\newcommand{\beqy}{\begin{eqnarray}}
\newcommand{\eeqy}{\end{eqnarray}}
\newcommand{\bitm}{\begin{itemize}}
\newcommand{\eitm}{\end{itemize}}

\begin{document}

\title{Multiple solutions in extracting physics information from
experimental data}
\author{C.~Z.~Yuan\footnote{yuancz@ihep.ac.cn},
X.~H.~Mo\footnote{moxh@ihep.ac.cn}, P.~Wang\footnote{wangp@ihep.ac.cn} \\
{\small Institute of High Energy Physics, Chinese Academy of Sciences, 
Beijing 100049, China} }

\date{}
\maketitle
\begin{center}
\begin{minipage}{15cm}
{\small
{\bf Abstract} \hskip 0.25cm
Multiple solutions exist in various experimental situations
whenever the sum of several amplitudes is used to fit the
experimentally measured distributions, such as the cross section,
the mass spectrum, or the angular distribution. We show a few
examples where multiple solutions were found, while only one
solution was reported in the publications. Since there is no
existing rules found in choosing any one of these solutions as the
physics one, we propose a simple rule which agrees with what have
been adopted in previous literatures: the solution corresponding
to the minimal magnitudes of the amplitudes must be the physical
solution. We suggest test this rule in the future experiments.

 }
\end{minipage}
\end{center}

\section{Introduction}

In quantum mechanics, a physics observable is proportional to the
modulus of the amplitude squared. In case of more than one
amplitude contributing to a process, they are summed to obtain the
total amplitude (generally there are relative phases between these
amplitudes), and thus one generally has contribution from
interference term to the physics observable. It is simple to
predict a physics observable when the amplitudes and the relative
phases between them are known, since there is no ambiguity in this
procedure.

However, in many circumstances, the experimental quantities are
measured, and from which we extract the information on the
amplitudes. As there is a square operation between the observable
and the amplitudes, we would expect multiple solutions in solving
the equation from a pure mathematics point of view. Then we face a
common problem: which solution is the physics one.

In this Letter, we present a few examples where two-solutions were
reported in fitting with the coherent sum of two amplitudes; and
revisit a few cases where only one solution was reported and we
find the other distinctive solution in fitting with two
amplitudes; we also discuss more complicated situations where more
than two amplitudes are used to fit the data. Finally, we propose
a conjecture on how to choose from multiple solutions the physics
one.

\section{Examples with two-solutions reported}

\begin{figure}[htb]
\centerline{\psfig{file=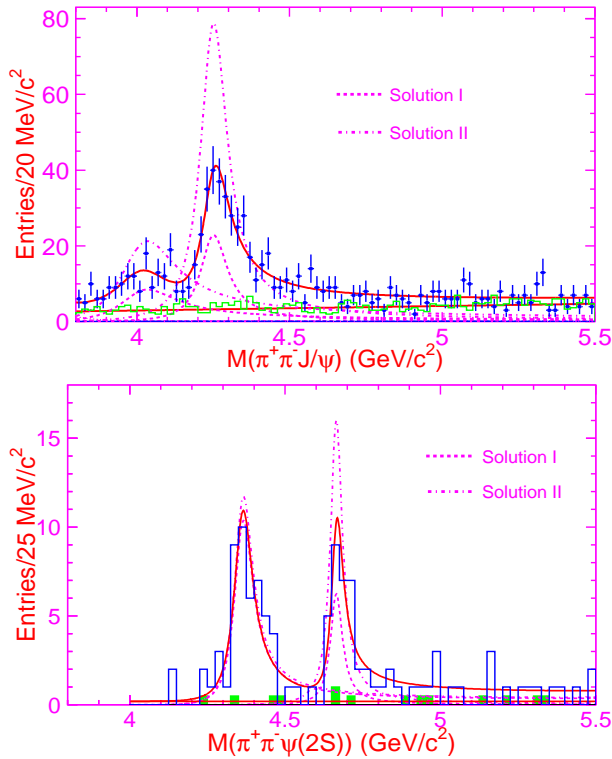,width=8cm,height=10cm}}
\caption{The $\pi^+\pi^-J/\psi$ (upper) and $\pi^+\pi^-\psi(2S)$
(lower) invariant mass distributions and the best fit with two
coherent resonances together with a background term. The data are
from Belle~\cite{belley,belle_pppsp}.} \label{mass}
\end{figure}

A few recent examples reporting multiple solutions are in the
study of the so-called $Y$ states via initial state radiation
($ISR$) by the Belle experiment. Figure~\ref{mass} shows the
invariant mass distributions of $\pi^+\pi^-J/\psi$ and
$\pi^+\pi^-\psi(2S)$ after all the selection in Belle
data~\cite{belley,belle_pppsp}, together with a fit with two
coherent resonant terms and an incoherent background term.
Table~\ref{tab1} shows the fit results, including the $Y(4008)$
and $Y(4260)$ from the $\pi^+\pi^-J/\psi$ mode, and the $Y(4360)$
and $Y(4660)$ from the $\pi^+\pi^-\psi(2S)$ mode. It should be
noted that in both channels there are two solutions of exactly the
same goodness-of-the-fit, with exactly the same mass and width for
the resonances but with very different coupling to $e^+e^-$ pair
($\Gamma_{e^+e^-}$). Instead of choosing one from the two
solutions, Belle reported both in their publications, however,
Particle Data Group (PDG) quoted sometimes only one of the
solutions in averaging with those from other
experiments~\cite{PDG}. Such a treatment of data is rather
suspicious, since the solutions picked up randomly from different
experiment may have distinctive features and should not be
averaged together.

\begin{table}[hbt]
\caption{\label{tab1} Fit results of the $\pi^+\pi^-J/\psi$ and
$\pi^+\pi^-\psi(2S)$ invariant mass spectra. The first errors are
statistical and the second systematic. $M$, $\Gamma_{\rm tot}$,
and ${\cal B}\cdot \Gamma_{e^+e^-}$ are the mass (in MeV), total
width (in MeV), product of the branching fraction to hadronic mode
and the $e^+e^-$ partial width (in eV), respectively. $\phi$ is
the relative phase between the two resonances (in degrees).}
\renewcommand{\arraystretch}{1.1}
\begin{center}
{
\begin{tabular}{ccc}
\hline\hline
  Parameters & Solution I & Solution II \\
  \hline
  $M(Y(4008))$            & \multicolumn{2}{c}{$4008\pm 40^{+114}_{-28}$}  \\
  $\Gamma_{\rm tot}(Y(4008))$   & \multicolumn{2}{c}{$ 226\pm 44\pm 87$}  \\
  ${\cal B}\cdot \Gamma_{e^+e^-}(Y(4008))$
                  & $5.0\pm 1.4^{+6.1}_{-0.9}$ & $12.4\pm 2.4^{+14.8}_{-1.1}$  \\
  $M(Y(4260))$            & \multicolumn{2}{c}{$4247\pm 12^{+17}_{-32}$} \\
  $\Gamma_{\rm tot}(Y(4260))$   & \multicolumn{2}{c}{$ 108\pm 19\pm 10$} \\
  ${\cal B}\cdot \Gamma_{e^+e^-}(Y(4260))$
                  & $6.0\pm 1.2^{+4.7}_{-0.5}$ & $20.6\pm 2.3^{+9.1}_{-1.7}$ \\
  $\phi$          & $12\pm 29^{+7}_{-98}$ & $-111\pm 7^{+28}_{-31}$
  \\\hline
    $M(Y(4360))$            & \multicolumn{2}{c}{$4361\pm 9\pm 9$} \\
  $\Gamma_{\rm tot}(Y(4360))$   & \multicolumn{2}{c}{$74\pm 15\pm 10$} \\
  ${\cal B}\cdot \Gamma_{e^+e^-}(Y(4360))$
                  & $10.4\pm 1.7\pm 1.5$ & $11.8\pm 1.8\pm 1.4$ \\
  $M(Y(4660))$            & \multicolumn{2}{c}{$4664\pm 11\pm 5$} \\
  $\Gamma_{\rm tot}(Y(4660))$   & \multicolumn{2}{c}{$48\pm 15\pm 3$} \\
  ${\cal B}\cdot \Gamma_{e^+e^-}(Y(4660))$
                  & $3.0\pm 0.9\pm 0.3$ & $7.6\pm 1.8\pm 0.8$ \\
  $\phi$           & $39\pm 30\pm 22$ & $-79\pm 17\pm 20$ \\
  \hline\hline
\end{tabular}}
\end{center}
\end{table}

Another example is the study of the decay dynamics of $\etap\to
\gamma\pp$ mode. When the $\pp$ invariant mass distribution is
fitted with coherent sum of the $\rho$ resonance and a contact
term, it is found that there are two solutions of equal
goodness-of-the-fit. One solution corresponds to constructive
interference between the two amplitudes while the other
destructive interference~\cite{etap_chenhx}. Similar study in
previous analyses only reported one solution~\cite{benayoun} which
corresponds to the solution with constructive interference in the
BES data~\cite{etap_chenhx}.

\section{Examples with one-solution reported}

As the examples shown in the previous section are basically a fit
to an $s$-dependent distribution with two coherent amplitudes, one
would expect that there would be in general two solutions in such
a circumstance. In the literatures, there are many such kinds of
fits, especially in the low energy $\EE$ annihilation experiments,
where the cross section of $\EE\to \hbox{hadrons}$ is
parameterized as the coherent sum of the amplitudes of the vector
mesons. We show two typical examples below where only one solution
was reported, and we redo the fit to obtain the other solution
from the data.

\subsection{Branching fraction of $\phi\to \omega\piz$}

The most precise data on $\EE\to \omega\piz$ near the $\phi$
resonance were reported by KLOE~\cite{kloe_omegapi} with both
$\omega\to \pp\piz$ and $\omega\to \gamma\piz$. The cross section
as a function of $\sqrt{s}$ is parameterized as
\[
\sigma(\sqrt{s}) = \sigma_{nr}(\sqrt{s})
\cdot\left|1-Z\frac{M_{\phi}\Gamma_{\phi}}{D_{\phi}(\sqrt{s})}\right|^2
\]
in the KLOE paper~\cite{kloe_omegapi}, where
$\sigma_{nr}(\sqrt{s}) = \sigma_{0} + \sigma' (\sqrt{s} - M_\phi)$
is the bare cross section for the non-resonant process,
parameterized as a linear function of $\sqrt{s}$; $Z$ is the
interference parameter, while $M_{\phi}$, $\Gamma_{\phi}$ and
$D_{\phi}=M_{\phi}^2-s-iM_{\phi}\Gamma_{\phi}$ are the mass, the
width, and the inverse propagator of the $\phi$ meson,
respectively.

\begin{figure}[htb]
\begin{minipage}{7cm}
\centerline{\psfig{file=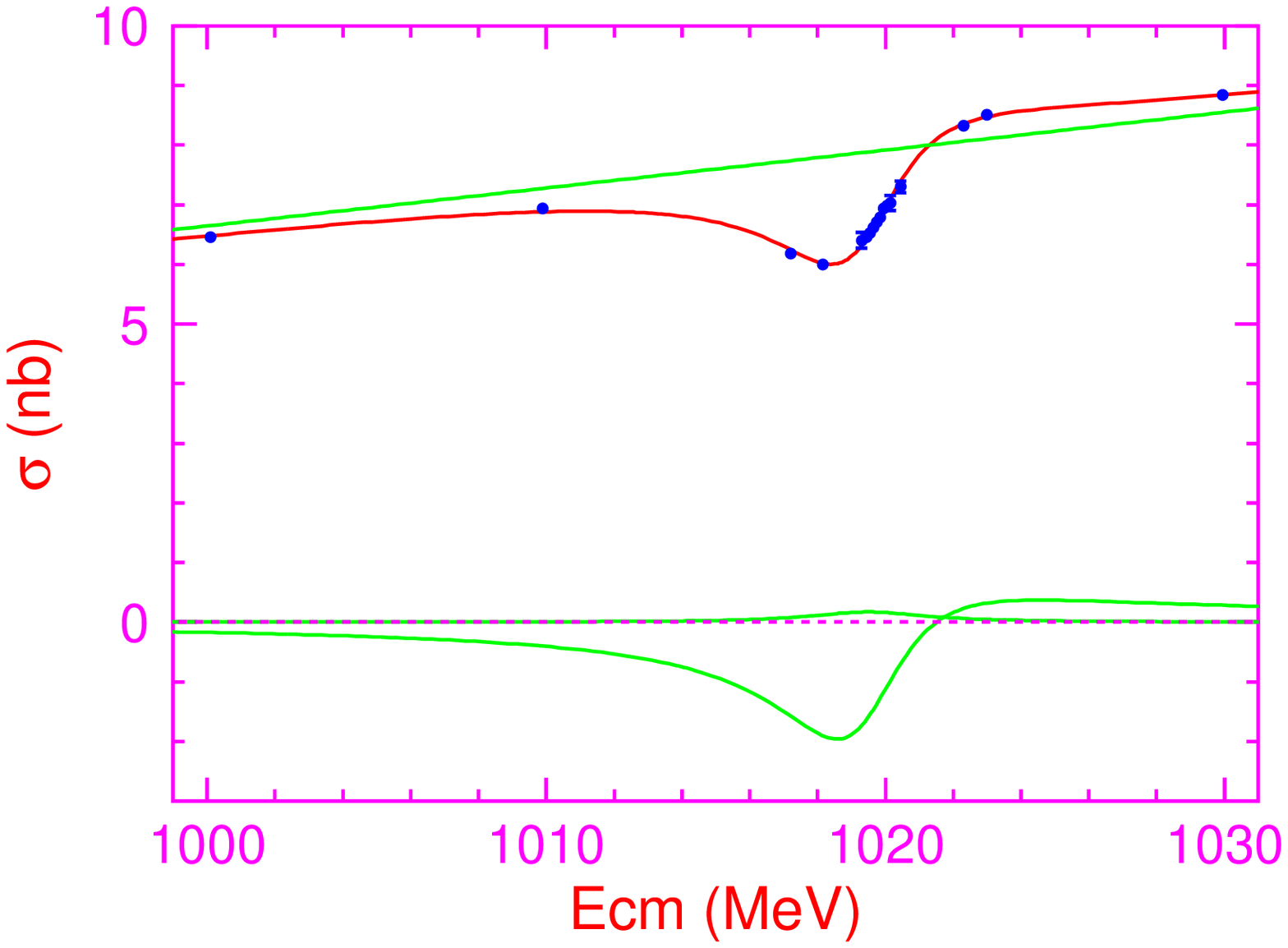,width=7cm,height=7cm,angle=-90}}
\end{minipage}
\begin{minipage}{7cm}
\centerline{\psfig{file=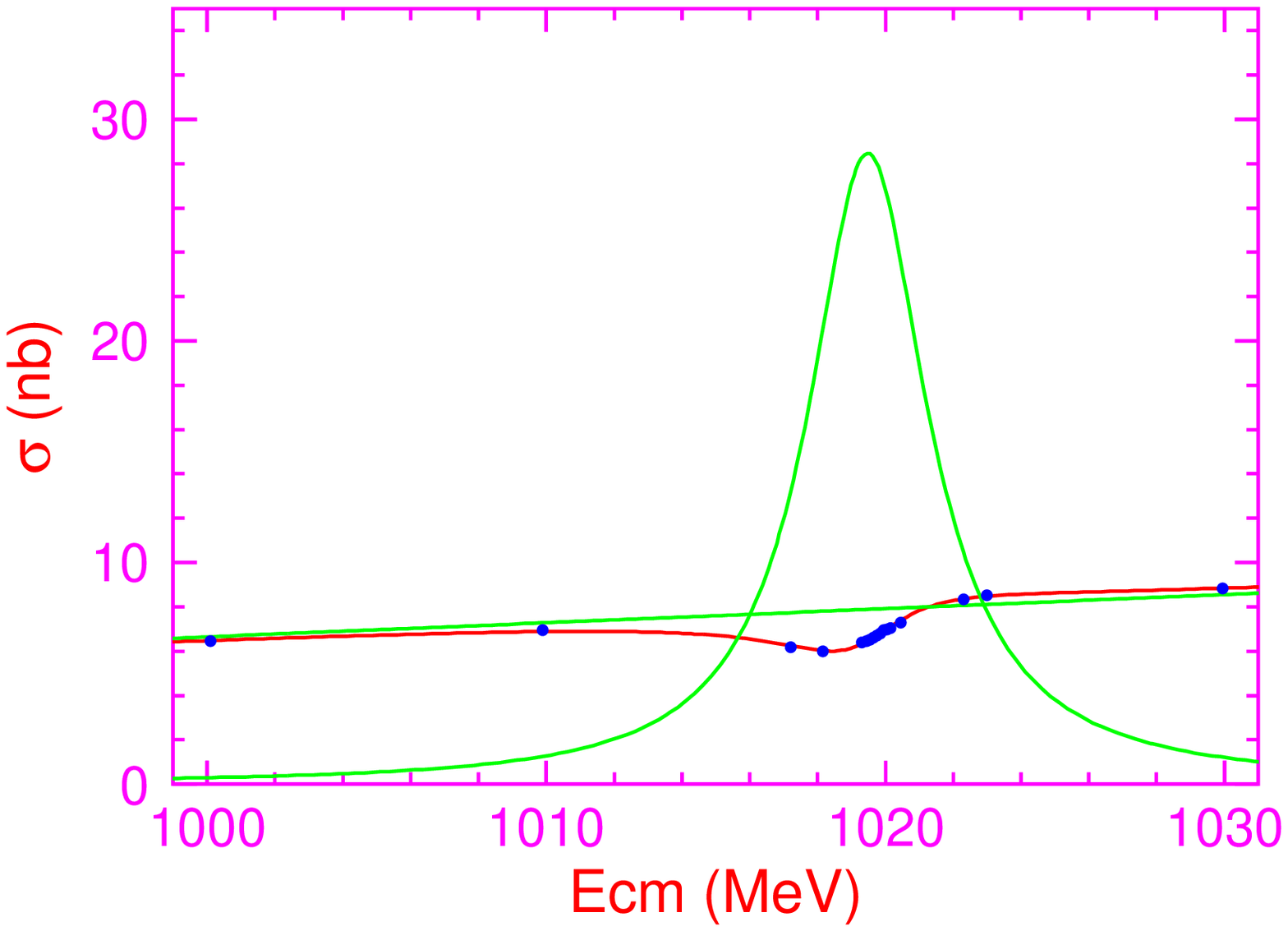,width=7cm,height=7cm,angle=-90}}
\end{minipage}
\caption{Fit to the $\EE\to \omega\piz$ cross sections as a
function of center-of-mass energy. Data points are shown with dots
with error bars. Top plot is for solution~I: at
$\sqrt{s}=1.019$~GeV, from top to bottom, the solid curves are
continuum, best fit, resonance, and interference terms. Bottom
plot is for Solution~II: at $\sqrt{s}=1.019$~GeV, from top to
bottom, the solid curves are resonance, continuum, and best fit,
the interference term is not shown.} \label{fit-omegapi}
\end{figure}

We take the KLOE data from Table~I of Ref.~\cite{kloe_omegapi}
with $\omega\to \pp\piz$ and fit with the same parametrization
given above. In our fit, the Born-order cross section is
calculated as $\sigma_{\rm vis}/\delta_{\rm rad}$, and only
statistical errors are considered in the $\chi^2$ construction.
Table~\ref{omegapi} shows the results from our fit, where two
solutions are found with the same fit quality. We can see that the
parameters from Solution~I are quite similar to those listed in
Table~II of Ref.~\cite{kloe_omegapi}, the slight difference is due
to the simplified procedure for the Born-order cross section
calculation in our fit. It is found that Solution~II has a much
larger resonant amplitude, and the resulting branching fraction of
the isospin violating process $\phi\to \omega\piz$ is at per mille
level, about two orders of magnitude higher than Solution~I,
namely, the solution reported in the original
work~\cite{kloe_omegapi}. Figure~\ref{fit-omegapi} shows the fit
results and the contribution of each component in the fit.
Similarly, we try a fit to the cross sections measured with
$\omega\to \gamma \piz$, there are also two solutions obtained
from the fit and the results are in good agreement with those from
$\omega\to \pp\piz$.

\begin{table}[htb]
  \caption{Results from fits to the $\EE\to \omega\piz$
cross sections measured with $\omega\to \pp\piz$.}
  \label{omegapi}
  \renewcommand{\arraystretch}{1.1} 
  \begin{center}
    \begin{tabular}{ccc}
      \hline\hline
      Parameter & Solution I & Solution II \\\hline
        $\sigma_0$ [nb]   & $ 7.88 \pm 0.04  $ & $ 7.88  \pm 0.08 $\\
        $\Re (Z)$        & $ 0.106 \pm 0.004 $ & $ 0.106  \pm 0.006 $\\
        $\Im (Z)$        & $-0.103 \pm 0.003 $ & $-1.90  \pm 0.006 $\\
        $\sigma'$ [nb/MeV]& $ 0.064 \pm 0.002 $ & $ 0.064 \pm 0.006$\\
        $\BR(\phi\to \omega\piz)$ & $4.61\times 10^{-5}$
        & $7.62\times 10^{-3}$ \\
        \hline\hline
      \end{tabular}
  \end{center}
\end{table}

\subsection{$\fpi$ and the $\rho$-$\omega$ mixing}

Even before the discovery of the $\tau$ lepton, Tsai calculated
the branching fractions of such a heavy lepton decaying into
vector final states by the conserved vector current (CVC)
hypothesis using $\EE\to \pp$ data measured in $\EE$ annihilation
experiments~\cite{tsai}. This calculation, referred to as $\BCVC$,
was tested ever since the discovery of the $\tau$ lepton. As the
precisions of both $\EE$ annihilation and $\tau$ decay experiments
improved significantly in the past three decades, the test has
reached the precision of better than 1\% level, and the
discrepancy of $2\sigma$ level between $\BCVC$ and $\Btau$ is
observed~\cite{md_tau_2009}. Many theoretical efforts have been
made to understand this discrepancy, including a better
understanding of the isospin violation correction and so on.

One source of the isospin breaking correction is the
$\rho$-$\omega$ mixing effect in the $\EE$ annihilation data which
is absent in the $\tau$ decays. The mixing effect on $\BCVC$ is
estimated by fitting the $\EE\to \pp$ data in the vector meson
dominance (VMD) model (including $\rho$ and its excited states and
$\omega$) and subtract the $\omega$ contribution by setting its
amplitude to zero. This has been carried out in many previous
analyses~\cite{md_tau_2009,maltman,cmd2}, but in all cases only
one solution was reported. As a heuristic example, we only use the
CMD2 data and follow the fit described in the CMD2
paper~\cite{cmd298}, namely, with the
GS-parametrization~\cite{gs68} of the $\fpi$:
\begin{eqnarray*}
F_{\pi}^{\rm GS}(s) & = & \left.\frac{
       \left.{\rm BW}_\rho^{\rm GS}(s)
\left[1 + \delta{\frac{s}{m_{\omega}^2}} P_\omega (s)\right] +
  \beta {\rm BW}_{\rho^\prime}^{\rm
  GS}(s)\right.}{1+\beta}\right.,
\end{eqnarray*}
where
\begin{eqnarray*}
{\rm BW}_V^{\rm GS}(s)  & = &
   \frac {m_V^2 ( 1+d\cdot \Gamma_V/m_V )}
   { m_V^2-s+f(s)- im_V\Gamma_V(s)}, \\
P_\omega (s) & = & \frac{m_{\omega}^2}
     {m_\omega^2-s- i m_{\omega}\Gamma_{\omega}}.
\end{eqnarray*}
The definitions of all the quantities can be found in
Refs.~\cite{cmd2,gs68}. Here, complex numbers $\delta$ and
$\beta$, as well as the mass and width of the $\rho$, are fit
parameters. The masses and widthes of $\omega$ and $\rho^\prime$
are fixed to their PDG values~\cite{PDG}, and the phase of $\beta$
is fixed to $180^\circ$.

\begin{figure}[htb]
\begin{minipage}{7cm}
\centerline{\psfig{file=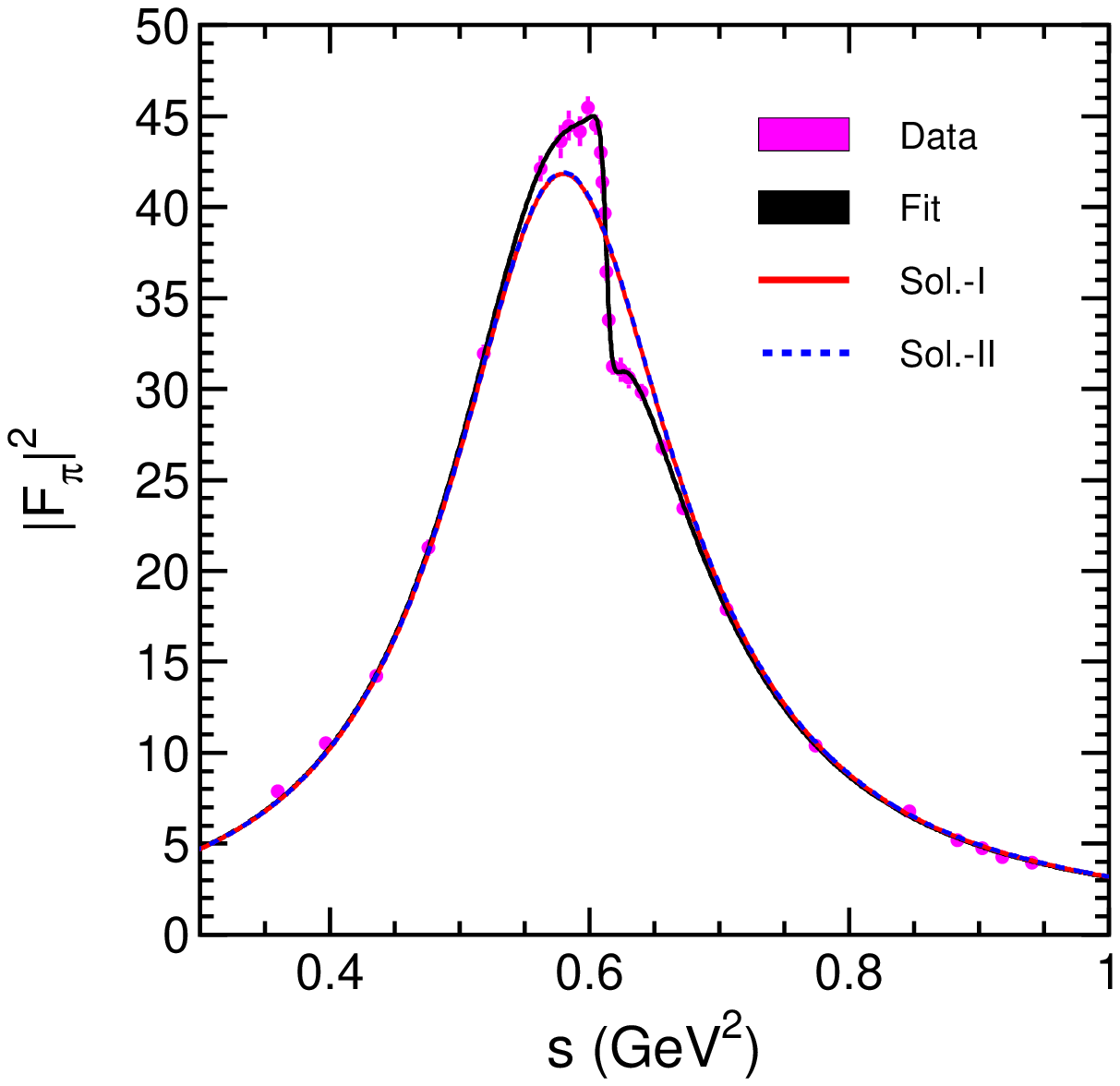,width=7cm,height=7cm}}
\end{minipage}
\begin{minipage}{7cm}
\centerline{\psfig{file=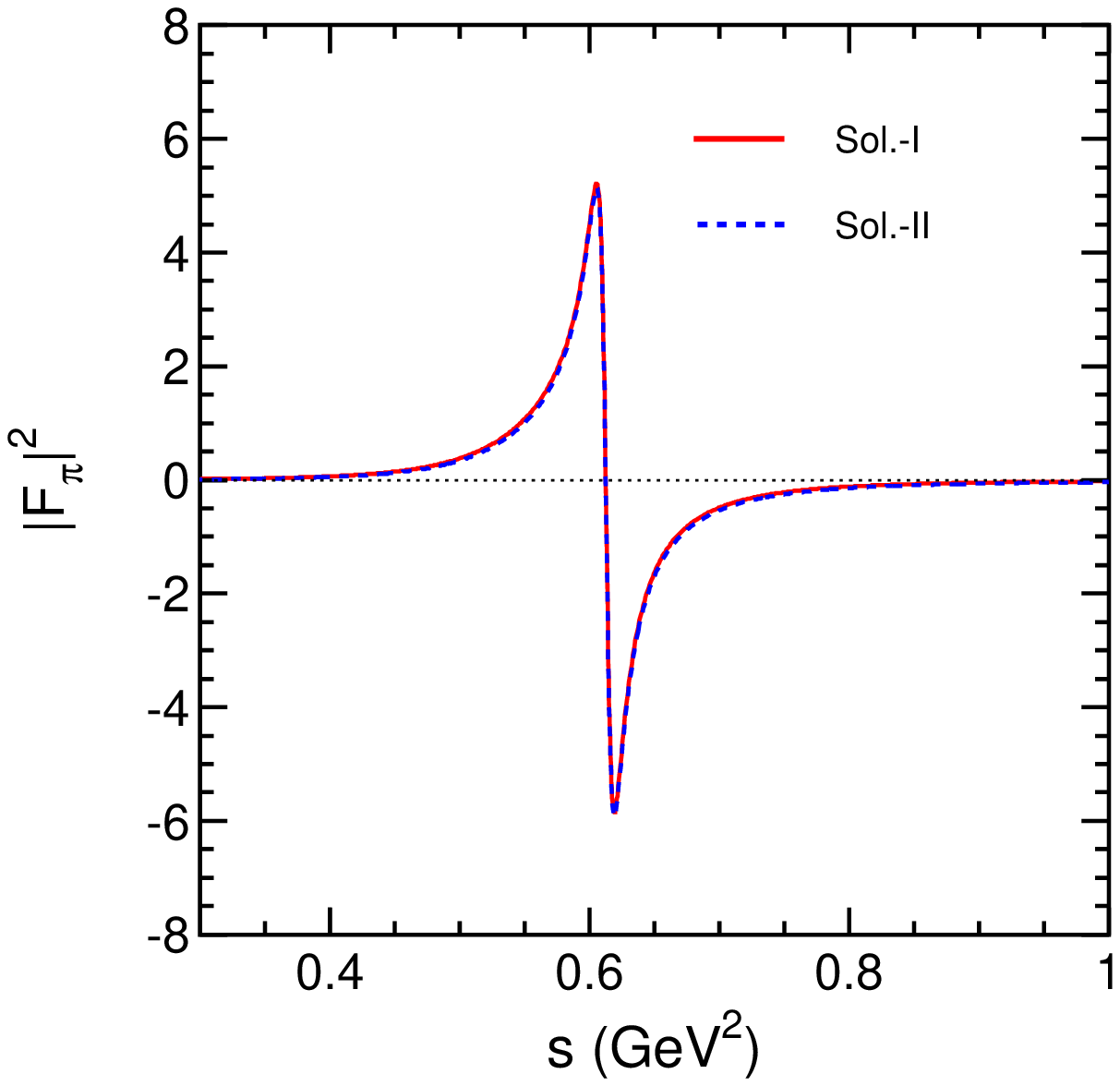,width=7cm,height=7cm}}
\end{minipage}
\caption{Fit to the $\EE\to \pp$ form factors below $s=1$~GeV$^2$
measured at CMD2~\cite{cmd2}. Top: along the data points (dots
with error bars) is the best fit. The $I=1$ part of the fit is
shown for the two solutions. Bottom: comparison of the
$\rho$-$\omega$ interference part in two solutions.}
\label{cmd2-fit}
\end{figure}

We found two solutions in our fit, as shown in Fig.~\ref{cmd2-fit}
and Table~\ref{cmd2-results}. It is clear that Solution~I is in
good agreement with the results reported by CMD2
experiment~\cite{cmd298}, and the resulting $\rho$-$\omega$ mixing
corrections to $\BCVC$ and $a_\mu$ are in agreement with the
results in Refs.~\cite{md_tau_2009,maltman}. However, the relative
strength between $\omega$ and $\rho$ is different in the other
solution, and the corrections to $\BCVC$ and $a_\mu$ are different
too, as shown in Table~\ref{cmd2-results}.

\begin{table}[htb]
  \caption{Results from fits to the $\EE\to \pp$
form factors measured at CMD2~\cite{cmd2}.}
  \label{cmd2-results}
  \renewcommand{\arraystretch}{1.1} 
  {\small
  \begin{center}
    \begin{tabular}{ccc|c}
      \hline\hline
      Parameter & Solution I & Solution II & Davier~\cite{md_tau_2009}\\\hline
        $m_\rho$ [MeV]      & \multicolumn{2}{c|}{$775.9\pm 0.5$} & --\\
        $\Gamma_\rho$ [MeV] & \multicolumn{2}{c|}{$146.0\pm 0.8$} & --\\
        $ |\delta|\;\; [\times 10^{-3}]$   & $1.62\pm 0.06$
                                                & $21.97\pm 0.04$ & --\\
        $ \phi_\delta \;\;[^\circ]$ & $10.1\pm 1.4$ & $86.56\pm 0.17$ & --\\
        $ |\beta| $    & \multicolumn{2}{c|}{$0.086\pm 0.004$} & --\\
        \hline
        $\Delta\BR^{\rm mixing}\;\; [\%]$ & $-0.03\pm 0.01$ & $+0.04\pm 0.01$
                             & $-0.01\pm 0.01$\\
        $\Delta a_{\mu}^{\rm mixing}\;\; [10^{-10}]$ & $+2.5\pm 0.2$ & $+1.6\pm 0.2$
                             & $+2.80\pm 0.19$\\
        \hline\hline
      \end{tabular}
  \end{center}}
\end{table}

As the amplitude of the $\omega$ term in Solution~II is more than
an order of magnitude higher than that in Solution~I, we would
expect a branching fraction of $\omega\to \pp$ much larger than
that reported in previous analyses: about 70\% in Solution~II
versus 1.5\% in Solution~I. It seems Solution~II is in
contradiction with our expectation that the isospin breaking decay
$\omega\to \pp$ should be much smaller than the isospin conserving
decay $\omega\to \pp\piz$. However, it is obvious that we are used
to the small $\omega\to \pp$ branching fraction because we have
never thought about the possibility of the existence of the second
solution in the fit.


\section{Examples with more than two amplitudes}

The examples shown above are described by the sum of two
amplitudes with a free relative strength and relative phase. In
the case of more than two amplitudes contributing to the
distribution, the number of solutions is $2^{n-1}$, where $n$ is
the number of free amplitudes.

In the fit to the $\EE\to \phi\pp$ cross
sections~\cite{belle_y2175}, incoherent and coherent sums of two
amplitudes (the $\phi(1680)$ and $Y(2175)$) are tested, while in
estimating the significance of a third resonance ($X(2400)$),
Belle also tested the possibility of the coherent sum of three
amplitudes. This procedure is repeated with combined
BaBar~\cite{babar_y2175} and Belle data~\cite{belle_y2175} in
Ref.~\cite{shencp_y2175}. In this latter case, four solutions are
found with the same masses and widthes for the resonances, but
with different coupling constants.

The same phenomenon was also shown in the fit to the combined
BaBar~\cite{babar_pppsp} and Belle~\cite{belle_pppsp} data on
$\EE\to \pppsp$~\cite{liuzq}, where two pairs of solutions were
found when fitting with the coherent sum of the $Y(4360)$,
$Y(4660)$, and $Y(4260)$.

\section{Partial wave analysis}

All the examples discussed so far are one-dimensional problems. In
the case of fitting a multi-dimensional distribution, the same
multi-solution problem also exists. The partial wave analysis
(PWA) and the Dalitz plot analysis techniques used widely in
hadron physics potentially have such a problem.

In a recent BES analysis of the $\psp\to \ppjpsi$
process~\cite{bes_ppjpsi}, the $\pp$ system is fitted with
contributions from $\sigma$, a contact term, and a possible tensor
amplitude $f_2$. It is curious to us why the two very similar models
shown in Figure~4 of Ref.~\cite{bes_ppjpsi} give very different
fractions of the $\sigma$ term. The aforementioned expositions lead
us to suspect that two different solutions are found in the two
models, one with constructive interference and the other with strong
destructive interference. The authors may try to excavate the other
solutions in each of those models used in fitting the experimental
data. Fortunately, according to the study above, we found that all
different solutions result in identical line shape of the resonance
(the same mass and width), so the resonant parameters should be
correct even if the branching fractions obtained from these analyses
are not reliable.

A general partial wave analysis may have many amplitudes and all
of them are added coherently to fit the data distributions. In
principle, there are multiple solutions, although we cannot judge
how many fold of ambiguity. In some circumstances, when more than
one solutions are found, a simple average is taken to be the best
estimation of the parameters, this is obviously non-physical,
since any one of the solutions is a good description of the data
but the average is generally not; while in some other cases, this
is just overlooked, which makes the comparison between experiments
looks very strange, since one compares two different solutions
from two experiments which are different by definition. Herein,
great care must be taken in treating the magnitudes of the
amplitudes in extracting them from multi-dimensional distributions
with the coherent sum of many amplitudes.

\section{Minimal amplitude conjecture}

In preceding sections, we examined a few cases where amplitudes
are extracted from experimentally observed distributions, and in
all the cases more than one solutions were found. We also
discussed the similar problem in PWA. All these indicate that this
problem actually is common in data analysis. Up to now, this
problem has not been taken into consideration seriously. Moreover,
no one has thought about how to pick up one solution from all the
possibilities, and there is indeed lack of solid physics argument
on how to choose one of the solutions as the physics one.

However, from the existing examples and our observation, we notice
that the solution with the smallest modulus of the amplitude
agrees with present expectation, such as in the $\phi\to
\omega\piz$ and the $\rho$-$\omega$ mixing cases. So it may not be
very surprising that the physics solution of any system
corresponds to the one where the modula of the amplitudes take the
minimal values among all the solutions. We call this selection
rule ``the minimal amplitude conjecture''.

Such a rule can be comprehended readily from a simple ``economic''
principle. One would expect that a solution in which the physics
observable is produced via two or more very large amplitudes with
strong destructive interference is less ``economic'' than the one
via small amplitudes but with constructive interference. In the
latter case, we always find maximum constructive interference
among amplitudes. With this economic ``minimal amplitude
conjecture'', we can determine the unique physics solutions of all
the aforementioned examples. This will make the life much simpler.

\section{Conclusion}

In this Letter, we put forth two important problems: multiple
solutions in fitting experimental data and the selection of the
physics solution among many possible solutions. As a matter of
fact, more and more cases with multiple solutions appear in recent
data analyses and even more may have been overlooked previously.
Here, we strongly suggest that all possible solutions be found out
and reported in the future analyses, in order to avoid a babel of
arguments and misleading theoretical deductions.

When confronting with multiple solutions, one has to make a
choice. We propose a ``minimal amplitude conjecture'': the
solution with minimal modula of the amplitudes is the physics one.
Such a rule ensures a unique solution is singled out from multiple
solutions. This conjecture could be treated as a principle which
should be tested experimentally by the future data analyses.

\section*{Acknowledgements}

This work is supported in part by the National Natural Science
Foundation of China (10775412, 10825524, 10935008), the Instrument
Developing Project of the Chinese Academy of Sciences (YZ200713),
Major State Basic Research Development Program (2009CB825203,
2009CB825206), and Knowledge Innovation Project of the Chinese
Academy of Sciences (KJCX2-YW-N29).



\end{document}